# Simulation of adaptive feedforward control for magnetic alloy cavity

Xiang Li*, Yang Liu

Institute of High Energy Physics, Beijing, China

Spallation Neutron Source Science Center, Dongguan, Guangdong, China

Abstract

The upgrade plan of the China Spallation Neutron Source aims to enhance the beam power from 100 kW to 500 kW. To achieve this, the plan involves incorporating three new magnetic alloy cavities while maintaining the existing system to enable double harmonic acceleration. As a consequence of the increased current intensity, the beam loading effect will be significantly amplified, presenting a considerable challenge for the low-level RF control system of the magnetic alloy cavity. To address this challenge, an adaptive feedforward algorithm has been developed to enable optimal control. In addition, comprehensive simulations of the algorithm have been successfully conducted to validate its.

Keywords: magnetic alloy cavity, adaptive feedforward algorithm

Ⅰ. Introduction

The China Spallation Neutron Source (CSNS) comprises an 80 MeV low-energy H- linac, a 1.6 GeV rapid cycling synchrotron, and a target station. In subsequent upgrade plans, the second phase of the CSNS will elevate the beam power from 100 kW to 500 kW[1]. The rapid cycling synchrotron RF system will incorporate three magnetic alloy cavities to supply fundamental and second harmonics to mitigate the space charge effects during injection and the early stages of acceleration[2]. The relevant RF parameters are shown in Table 1.

With the enhancement of beam power, the beam loading effects in the RCS will be notably intensified. This poses significant challenges for the control system. Since the broadband magnetic alloy cavity do not require tuning, direct feedback becomes challenging to implement. The power source system's nonlinearity at high power levels also constrains the utilization of beam feedforward[3]. To address these challenges, we have developed an adaptive feedforward with the aim of improving system control precision under the conditions of strong beam loading effects and high power system nonlinearity.

Table 1. RF parameters comparison between CSNS and CSNS-II.

|  | CSNS | CSNS-II |
| --- | --- | --- |
| repetition rate / Hz | 25 | |
| beam power / kW | 100 | 500 |
| RF frequency / MHz | 1.02-2.44 | 1.71-2.44 |
| circulating current /A | 1.5-3.6 | 12.6-18 |
| number of RF cavities | 8 (fundamental harmonic) | 8(fundamental harmonic) 3(2nd harmonic) |

## Ⅱ. Algorithm studied by Simulink

### 1. System Modeling

Given that the magnetic alloy cavity is a broadband system, modeling the system using transfer functions is quite challenging. Therefore, based on the measured impedance data, we modeled the magnetic alloy cavity using Simulink shown in Fig.1. And the flow and operation of the adaptive feedforward module was established according to Fig.2.

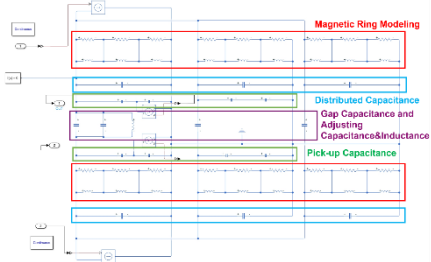

Fig.1 The magnetic alloy cavity modeling

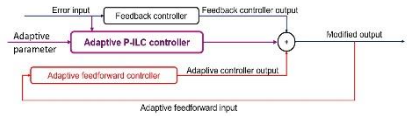

Fig.2 The flow and operation of the adaptive feedforward module

The magnetic alloy cavity modeling includes 45 pF gap capacitance, 25 μH Q-adjusting inductance, 50 pF tuning capacitance, 348.27 Ω magnetic ring impedance, 24.3158 μH magnetic ring inductance, 135 pF distributed capacitance, 25pF and 50 nF sampling capacitance combination.

The adaptive feedforward algorithm is based on the Iterative Learning Control (ILC) algorithm. As the magnetic alloy cavity system is broadband and operates in a sweeping frequency state, the system gain differs at various working frequencies. The relationship between the drive signal and the reference signal can effectively reflect changes in system gain. Therefore, we augmented the adaptive feedforward based on the drive signal with an adaptive gain parameter P-ILC.

### 2. Feed-forward Signal Processing

By transmitting data from each iteration to the workspace, one can filter it within MATLAB's workspace. The choice of the filter is crucial. Based on the ILC convergence conditions[4][5], the H-infinity ($H^\infty$) can be calculated for the magnetic alloy cavity system, which indicates that the filter bandwidth needs to be below 5kHz to make the iteration result converge. The Fig.3 shows $H^\infty$ for various filters.

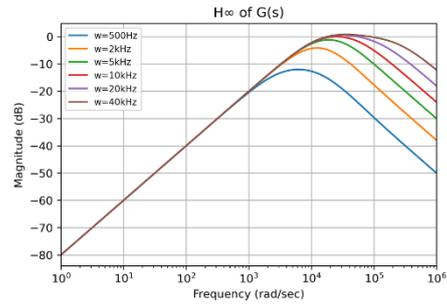

Fig.3 $H^\infty$ for various filters.

Using a conventional IIR filter introduces phase-lag, but using a zero-phase filter avoids this issue. We used a low-pass zero-phase filter with an 5kHz cutoff for the feedback signals. Subsequently, the processed feedforward signal is routed through the "from workspace" module.

### 3. Simulation Results

During the simulation, we experimented with different feedforward filter bandwidths. When the filter bandwidth exceeded 5kHz, oscillations occurred during the iterations. this is consistent with the theoretical results. However, when the bandwidth was set to 5kHz, the system remained stable. Based on theoretical analysis and practical simulations, setting the filter bandwidth to 5kHz proved to be feasible. The control error after taking adaptive feedforward for five iterations is shown Fig.4.

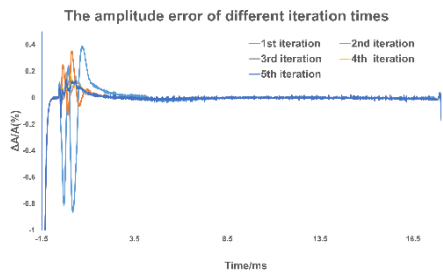

Fig4. The control error in 5 times iteration.

Comparative graphs depicting the control accuracy between the combined approach of adaptive feedforward[6] and adaptive P-ILC, , versus using just the adaptive feedforward or adaptive P-ILC are presented in Fig.5.

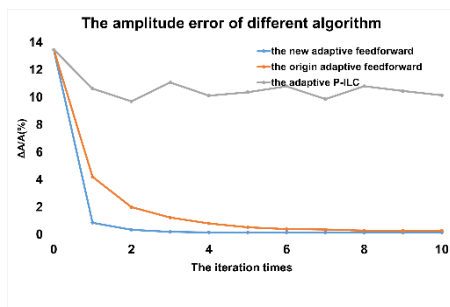

Fig5. The control accuracy of different algorithms.

Upon employing the new adaptive feedforward, the system's convergence speed significantly increased, and the control accuracy was also enhanced. After two times iteration, the new adaptive feedforward control effect is close to optimal. The origin adaptive feedforward needs 7 times iterations. And the maximum control error of the new adaptive feedforward after convergence is 1.4 ‰, while the maximum control error of the original adaptive feedforward through improved filters is twice that, so it can be seen that the new adaptive feedforward has a significant effect.

III．Conclusion and Future Outlook

The simulation results indicate that the use of the new adaptive feedforward has effectively improved the system's convergence speed and control accuracy. However, the simulation process cannot simulate the saturation distortion of the power source at high power and the beam loading effect. This optimization scheme still needs to be validated in the future and will be evaluated for the long-term performance of the improved feedforward control in the actual LLRF system. In addition, theoretical analysis will be conducted to study the impact of the new adaptive feedforward on control stability, convergence, and accuracy.